\documentclass[a4paper,12pt]{article}
\begin{document}

{\noindent \sf \large Analytical solution of the integral equation for partial wave \\
Coulomb $t$-matrices at excited-state energy } \\[.2in]
{\sf V. F Kharchenko}\\[.1in] 
{\footnotesize Bogolyubov Institute for Theoretical Physics,  
National Academy of Sciences \\of Ukraine, UA - 03143, Kyiv, Ukraine \\ [.1in]
E-mail: vkharchenko@bitp.kiev.ua} \\[.1in]
\noindent \small{ {\sf Abstract }  \\ 
{\small Starting from the integral representation of the three-dimensional 
Coulomb transition matrix elaborated by us formerly with the use of specific
symmetry of the interaction in a four-dimensional Euclidean space introduced by Fock, 
the possibility of the analytical solving of the integral equation for the partial   
wave transition matrices at the excited bound state energy has been studied.
New analytical expressions for the partial $s$-, $p$- and $d$-wave Coulomb 
$t$-matrices for like-charged particles and the expression for the partial 
$d$-wave $t$-matrix for unlike-charged particles at the energy of the first 
excited bound state have been derived.} \\ [.05in]

\noindent {\sf 1. Introduction} \\ 

\noindent The Coulomb transition matrix, which in the momentum space is a function of the 
initial and final relative momenta and the energy, provides all information about 
the system of two charged particles with the Coulomb interaction [1]. The knowledge 
of the Coulomb transition matrix with the momevta off the energy shell is necessary 
when studing properties of the few-body atomic and nuclear systems containing 
charged particles by using the Faddeev [2,3] and Faddeev-Yakubovskii [4] integral 
equations. The separation of the main Coulomb singularity and the regularization 
of the integral equations for the three-body system, which contains charged particles, 
has been performed by Vesselova [5,6] using the known Gorshkov's receipe for the system 
of two charged particles [7]. 

Presently, there are known several representations of the two-body Coulomb 
transition matrix [8-18]. The one-parameter integral representations for the 
Coulomb Green's function which take imediately into account the existing 
symmetry of the Coulomb system in Fock's four-dimensional Euclidean space [19] 
have been firstly obtained in the papers by Bratsev and Trifonov [10] and 
Schwinger [13]. The expressions for three-dimensional Coulomb transition matrix 
with the explicit removal of the singularities in the transfer-momentum variable 
and the energy have been derived in the papers [15] (for the negative values of the 
energy $E<0$) and [16] (for zeroth and positive energies $E \geq 0$). The 
possibility of obtaining the expressions for partial two-body Coulomb transition 
matrix at the energy of the ground bound state in the analytical form was 
investigated in previous papers [20] ( in the case of the attractive interaction) 
and [21] (in the case of the repulsive interaction ). 

This paper is devoted to the derivation of the analytical expressions for the 
partial wave Coulomb transition matrices at the energy of the first excited 
state of the two-body system. In Section 2 we begin the derivation leaning upon
the integral representation of the three-dimensional Coulomb $t$-matrix obtained 
in [15]. On this basis in Section 3 we deduce the general formula for the 
partial wave Coulomb matrix $\langle {\bf k} \mid t_l(E) \mid {\bf k}^{\prime}\rangle$  
at the energy $E<0$. In Section 4 we study the simplification of the expression 
for the partial wave Coulomb $t$-matrix at the energy of the first excited 
state of the two-body bound system. In Section 5 analytical expressions for the 
partial $s$-, $p$- and $d$-wave Coulomb transition matrices in the case of the 
repulsive interaction are derived. Section 6 is devoted to the derivation of the 
analytical expressions for the partial $d$-wave transition matrix in the case of the 
attractive Coulomb interaction. The discussion of the obtained results, concluding 
remarks on the performed work and the plans of the future studing are given in 
Section 7. \\ 

\noindent {\sf 2. Three-dimensional Coulomb transitions matrix at negative energy}\\ 

In the momentum space the three-dimensional Coulomb transition matrix 
$<{\bf k}|t(E)|{\bf k}^{\prime}>$, which describes the system of two charged particles, 
is given by the Lippmann-Schwinger equation
\begin{equation}
<{\bf k}|t(E)|{\bf k}^{\prime}>=\langle {\bf k} \mid v \mid {\bf k}^{\prime}\rangle + 
\int \frac{d{\bf k}^{\prime\prime}}{(2\pi)^3}\langle {\bf k} \mid v \mid {\bf k}^{\prime\prime}\rangle
\frac{1}{E-\frac{k^{\prime\prime 2}}{2\mu}} <{\bf k}^{\prime\prime}|t(E)|{\bf k}^{\prime}>\;\;,   
\end{equation}  
with the free term in the form of the matrix of the Coulomb interaction between the particles 
1 and 2 
\begin{equation}
\langle {\bf k} \mid v \mid {\bf k}^{\prime}\rangle = 
\frac{4\pi q_1 q_2}{\mid {\bf k}-{\bf k}^{\prime}\mid ^2}\;. 
\end{equation}
and the kernel, that is defined by the potential of interaction and the free propagator.
Here $q_1$ and $q_2$ are the charges of the particles that interact between themselves, 
$E$ is the energy of the relative motion of the particles, $\mu$ is the reduced mass of 
the particles, $k$ and $k^{\prime}$ are the variable relative momenta in the final and 
the initial states, $\hbar$ is reduced Planck's constant.

In the case of the relative energy $E=-\hbar^2\kappa^2/2\mu$ the three-dimensional 
Coulomb transition matrix has the form [15]
\begin{displaymath}
 <{\bf k}|t(E)|{\bf k}^{\prime}>=\frac{8\pi q_1 q_2 \kappa^2}{(k^2+\kappa^2)
(k^{\prime 2}+\kappa^2)\sin\omega}\left\{\cot \frac{\omega}{2}
-\pi\gamma \cos\gamma\omega
-\gamma \sin 2\gamma \omega \ln (\sin\frac{\omega}{2}) \right.
\end{displaymath}
\begin{equation}
 \left. + 2\pi\gamma\; c(\gamma)\; \cot \gamma\pi \sin\gamma\omega 
+ \gamma\cos\gamma\omega x_{\gamma}(\omega) + 2\gamma^2 \sin\gamma\omega
y_{\gamma}(\omega) \right\}\;,
\end{equation} 
where 
\begin{equation}
\gamma = \frac{\mu q_1 q_2}{\hbar^2 \kappa}\;,
\end{equation} 
is the known dimensionless Coulomb parameter (the Sommerfeld parameter).
The variable quantity $\omega$ in Eq.(3) denotes the angle between two 
4-dimensional vectors in the four-dimensional Euclidean space introduced 
by Fock [19], the three-dimensional vectors ${\bf k}$ and ${\bf k}^{\prime}$ 
lie in the hyperplane, which is the stereographic projection of the unit 
sphere. The variable $\omega$ is defined by the following expression
\begin{equation}
\sin^2\frac{\omega}{2} = \frac{{\kappa^2}\mid {\bf k} - {\bf k}^\prime \mid ^2}
{(k^2 + \kappa^2)(k^{\prime 2} + \kappa^2)}\;\; , \;\; 0\leq \omega \leq \pi\;\;.
\end{equation} 

The functions  $x_{\gamma}(\omega)$, $y_{\gamma}(\omega)$ and $c(\gamma)$ in 
(3) are given by 
\begin{displaymath}
x_{\gamma}(\omega)= \int_{0}^{\omega} d\varphi \;\sin \gamma\varphi\;
\cot \frac{\varphi}{2} \;, y_{\gamma}(\omega)= \int_{\omega}^{\pi} 
d\varphi \;\sin \gamma\varphi\;\ln\left(\sin \frac{\varphi}{2}\right)\;,
\end{displaymath}
\begin{equation}
c(\gamma)= \frac{1}{2} \left[ 1 - \frac{1}{\pi} x_{\gamma}(\pi)\right]\;.
\end{equation} 

The singularities in the energy are contained in the fourth term of the 
expression in the braces (3). They arise only in the case of the attractive 
Coulomb interaction potential (for oppositely charged particles with $q_1 q_2 <0$), 
when the Coulomb parameter $\gamma$ takes on the negative integer values ($\gamma 
= -n,\; n=1,2,3,...$) corresponding to the spectrum of the bound states of the 
two-particle system with the energies
\begin{equation}
E_n = - \frac{\mu (q_1 q_2)^2}{2 \hbar^2 n^2}\;, \qquad  n=1, 2, \cdots \;.
\end{equation}
The values $\kappa$ and the Coulomb parameter $\gamma$ which correspond to the 
energy (7) are equal to
\begin{equation}
\kappa_n = \frac{\sqrt{-2\mu E_n}}{\hbar} = \frac{\mu \mid q_1 q_2 \mid }
{\hbar^2 n}\;, \qquad \gamma_n = \frac{\mu q_1 q_2}{\hbar^2 \kappa_n} = 
\frac{q_1 q_2}{\mid q_1 q_2 \mid}n\;.
\end{equation}
For the energies of the bound states $E=E_n$ the Coulomb parameter $\gamma$ 
takes the integer values --- positive for the repulsive Coulomb interaction 
($q_1 q_2 >0,\; \gamma_n=n$) and negative for the attractive interaction
($q_1 q_2 <0,\; \gamma_n=-n$). Thus, taking into account that
\begin{equation}
x_n(\pi) = \pi\;,\qquad   x_{-n}(\pi) = - \pi\;,
\end{equation}
we find from the expression for $c(\gamma)$ (7)
\begin{equation}
c(n) = 0\;,\qquad   c(-n) = 1\;.
\end{equation}

In the case of the repulsive interaction ($\gamma > 0$) the expression for the 
three-dimensional Coulomb transition matrix (3) has no singularities at the 
energies of the bound states $E=E_n$. The indeterminateness in the fourth 
term in the braces of the type $\frac{0}{0}$ at $\gamma=n$ we evaluate according 
to l'Hospital rule
\begin{equation}
\left[ \frac{2\pi \gamma c(\gamma)}{\tan \gamma\pi}\right]_{\gamma \rightarrow n} 
= 2 n c^\prime (n)\equiv \rho_n\;,
\end{equation}
where
\begin{equation}
\rho_n = (-1)^n - 2 n \ln 2 - 2 n \sum_{m=1}^{n} \frac{(-1)^m}{m} \;.
\end{equation}
\\ 
\noindent {\sf 3. Partial wave transition matrices at the negative energy}\\ 

Developing the matrix elements of the Coulomb potential and the corresponding 
three-dimensional transition matrix in Legendre polinomials $P_l(x)$
\begin{equation}
\langle {\bf k} \mid v \mid {\bf k}^{\prime}\rangle = \sum_{l=0}^{\infty} (2l+1) 
v_l(k,k^{\prime}) P_l(\hat{{\bf k}}\cdot \hat{{\bf k}}^{\prime})\;,
\end{equation}
\begin{equation}
\langle {\bf k} \mid t(E) \mid {\bf k}^{\prime}\rangle = \sum_{l=0}^{\infty} (2l+1) 
t_l(k,k^{\prime};E) P_l(\hat{{\bf k}}\cdot \hat{{\bf k}}^{\prime})\;,
\end{equation}
where $\hat{{\bf k}}$ is the unit vector along the vector ${\bf k}$ and  
$\hat{{\bf k}}\cdot\hat{{\bf k}}^{\prime}=\cos \theta$, we write the one-dimensional 
integral equation for the partial wave transition matrix in the form 
\begin{equation}
t_l(k,k^{\prime};E) = v_l(k,k^{\prime})+
\int_{0}^{\infty} \frac{dk^{\prime\prime} {k^{\prime\prime}}^2}{2\pi^2}
v_l(k,k^{\prime\prime}) \frac{1}{E-\frac{{k^{\prime\prime}}^2}{2\mu}}
t_l(k^{\prime\prime},k^{\prime};E)\;\;, \\[3mm]  
\end{equation}  

The inhomogeneous term and the kernel of this equation are expressed 
the partial wave component of the Coulomb interaction potential
\begin{equation}
v_l(k,k^{\prime}) = \frac{1}{2} \int_{0}^{\pi} d\theta \;
\sin \theta \;P_l(\cos \theta)
\langle {\bf k} \mid v \mid {\bf k}^{\prime}\rangle \;,
\end{equation}
that according to (2) can be written in the form
\begin{equation}
v_l(k,k^{\prime}) = \frac{2\pi q_1 q_2}{k k^{\prime}} Q_l 
\left( \frac{k^2+{k^{\prime}}^2}{2 k k^{\prime}} \right)\;,
\end{equation}
where the function $Q_l(x)$ is the Legendre polynomial of the second kind [22].

According to the definition (14), the partial wave component of the Coulomb 
transition matrix is equal to
\begin{equation}
t_l(k,k^{\prime};E) = \frac{1}{2} \int_{0}^{\pi} d\theta \;\sin \theta \;
P_l(\cos \theta)\;\langle {\bf k} \mid t(E) \mid {\bf k}^{\prime}\rangle \;.
\end{equation}

Taking the relationship of the angle between two 4-dimensional vectors in the 
four-dimensional Fock's space $\omega$ and the angle $\theta$ between the vectors 
${\bf k}$ and ${\bf k}^{\prime}$, we find
\begin{equation}
\cos {\theta} = \frac{\xi}{\eta} - \frac{1}{\eta} {\sin}^2{\frac{\omega}{2}} = 
\frac{2\xi - 1 + \cos {\omega}}{2\eta}\;,   
\end{equation}  
where 
\begin{equation}
\xi = \frac{\kappa^2 (k^2 + {k^{\prime}}^2)}{(k^2 + \kappa^2)({k^{\prime}}^2 + 
\kappa^2)}\;,\quad \eta = \frac{2 \kappa^2 k k^{\prime}}{(k^2 + 
\kappa^2)({k^{\prime}}^2 + \kappa^2)}\;.
\end{equation}  

Passing in (18) from the integration over the angle $\theta$ to  the integration over 
the variable $\omega$, we write the expression (18) for the partial wave component of 
the Coulomb transition matrix in the form 
\begin{equation}
t_l(k,k^{\prime};E) = \frac{1}{4\eta} \int_{\omega_0}^{\omega_{\pi}} d\omega\; 
\sin \omega \; P_l \left( \frac{2\xi-1+ 
\cos \omega}{2\eta} \right)
\langle {\bf k} \mid t (E) \mid {\bf k}^{\prime}\rangle  \;,
\end{equation}
where the integration limits in (22) are determined by the expressions
\begin{equation}
\omega_0 = 2 \arcsin \sqrt{\xi - \eta}\;, \qquad \; \omega_{\pi} = 2 \arcsin \sqrt{\xi + \eta}\;,
\end{equation} 
in this case 
\begin{displaymath}
\sin \omega_0 = 2 \sqrt{\xi - \eta} \sqrt{1 - \xi + \eta}\;, \qquad \; 
\sin \omega_{\pi} = 2 \sqrt{\xi + \eta} \sqrt{1 - \xi - \eta}\;.
\end{displaymath}
\begin{equation}
\cos \omega_0 = 1 - 2 \xi + 2 \eta\;, \qquad \; \cos \omega_{\pi} = 1 - 2 \xi - 2 \eta\;,
\end{equation} 

Substituting the expression (3) for the three-dimensional transition matrix into (21), we 
obtain the formula for the partial wave Coulomb transition matrix $t_l(k,k^{\prime};E)$ 
at $E<0$: 
\begin{displaymath}
t_l(k,k^{\prime};E)= \frac{\pi q_1 q_2}{k k^{\prime}} \int_{\omega_0}^{\omega_{\pi}}
d\omega \;P_l \left( \frac{2\xi-1+ \cos \omega}{2\eta}\right) 
\left\{ \cot {\frac{\omega}{2}} \right.
\end{displaymath}
\begin{equation}
- \pi \gamma \; \cos \gamma \omega - \gamma \;\sin 2\gamma\omega \;\ln (\sin \frac{\omega}{2}) 
+ 2\pi \gamma\; c(\gamma)\; \cot \gamma\pi\; \sin \gamma\omega  
\end{equation} 
\begin{displaymath}
\left. + \gamma \;\cos \gamma \omega x_{\gamma}(\omega) + 2\gamma^2 \sin {\gamma\omega} 
y_{\gamma}(\omega)\right\} \; .
\end{displaymath}
\newpage
\noindent {\sf 4. Partial-wave Coulomb transition matrices at the energy of the excited state }\\ 

It is easy to see that the expression (24)for the partial wave Coulomb t-matrix is significantly 
simplified, if the energy $E$ is equal to the values of the bound states energies (7), 
$E=E_n$, at which the parameter $\gamma$ takes the integer values. (Note that in the case of 
the attractive interaction ($q_1 q_2<0$) this statement is relevant only for the partial-wave 
transition matrices with the orbital momenta $l\geq n$, which have not the pole singularities 
at these values of the energy.)

At $\gamma=\pm 1$, that corresponds to the bound ground state energy $E_1$, the functions (6)
have a simple form
\begin{equation}
x_1(\omega) =  \omega + \sin\omega \;,  \qquad y_1(\omega) = \cos^2 \frac{\omega}{2}\;+\;
2\sin^2 \frac{\omega}{2}\; \ln\sin\left(  \frac{\omega}{2} \right)\;,
\end{equation} 
in this case
\begin{displaymath}
x_{-1}(\omega) = - x_1(\omega)\;  \qquad  \qquad  y_{-1}(\omega) = - y_1(\omega)\;, 
\end{displaymath}
This ensures a possibility to obtain the analytical expression for the partial wave Coulomb 
transition matrices at the energy $E_1$ for both the attractive [20] and repulsive [21] Coulomb 
interactions.

In this paper we extend the above results for the partial wave transition matrices at the ground-
state energy to a case when the variable of the energy $E$ takes the value of the energy of the 
first excited state $E_2$. Then the functions $x_2(\omega)$ and $y_2(\omega)$ in the expression (24) 
take the form
\begin{equation}
x_2(\omega) = \omega +  2\sin\omega + \frac{1}{2} \sin 2\omega\;,  \qquad y_2(\omega) = 
- \cos^4 \frac{\omega}{2}\;-\;\sin^2 \omega\; \ln \sin\left(  \frac{\omega}{2} \right)\;,
\end{equation} 

In the case of the repulsive Coulomb interaction ($q_1 q_2>0$, inserting (26) into the formula 
for the partial wave Coulomb transition matrix (24) at $E=E_2$ ($\gamma=2$) and using (10) - (12), 
we obtain
\begin{displaymath}
t^r_l(k,k^{\prime};E)= \frac{\pi q_1 q_2}{k k^{\prime}} \int_{\omega_0}^{\omega_{\pi}}
d\omega \;P_l \left( \frac{2\xi-1+ \cos \omega}{2\eta}\right) 
\left\{ \cot {\frac{\omega}{2}} \right.
\end{displaymath}
\begin{equation}
- 2\pi \; \cos 2\omega +  2\omega \; \cos 2\omega - 4 \sin\omega 
\end{equation} 
\begin{displaymath}
\left. + (\rho_2 -3)\sin2\omega - 4 \sin 2\omega \ln \left( \frac{\omega}{2} \right)     
\right\} \; ,
\end{displaymath}
where according to (12)
\begin{equation}
\rho_2 = 3 - 4 \ln 2\;. 
\end{equation} 

In the case of the attractive interaction ($q_1 q_2<0$, the formula for the partial wave 
Coulomb transition matrix (24) at $E=E_2$ ($\gamma=-2$) takes the form
\begin{displaymath}
t^a_l(k,k^{\prime};E)= \frac{\pi q_1 q_2}{k k^{\prime}} \int_{\omega_0}^{\omega_{\pi}}
d\omega \;P_l \left( \frac{2\xi-1+ \cos \omega}{2\eta}\right) 
\left\{ \cot {\frac{\omega}{2}} \right.
\end{displaymath}
\begin{equation}
+ 2\pi \; \cos 2\omega - 2\omega \; \cos 2\omega - 4 \sin\omega 
\end{equation} 
\begin{displaymath}
\left. -3 \sin2\omega - 4 \sin 2\omega \ln \left( \frac{\omega}{2} \right)     
\right\} \; ,
\end{displaymath}

Herewith, it must be remembered that the singularities in the energy of the three-dimensional 
Coulomb transition matrix for the attractive interaction in (3) pass only to the partial wave 
transition matrices with the orbital momenta $l<n$. All other partial wave components of the 
transition matrix are nonsingular at $E=E_n$. This follows from the form of the fourth term 
in (24) which contains the coefficient
\begin{equation}
I_{\gamma l} \equiv  \int_{\omega_0}^{\omega_{\pi}}
d\omega \;\sin\gamma\omega \;P_l \left( \frac{2\xi-1+ \cos \omega}{2\eta}\right) 
\end{equation} 
before the singular term $\cot \gamma\pi$ . In the case $\gamma = -2$, which is under study, 
for example, we have
\begin{equation}
I_{-2 l} = -\frac{16}{3}\eta^2 \delta_{l 1} + 8\eta (2\xi - 1)\delta_{l 0}\;.  
\end{equation}  
In this way, all the partial wave transition matrices with the orbital momenta $l\geq 2$ have 
no pole singularities. \\ 

\noindent {\sf 5. Analytical expressions for the partial-wave transition matrices at the energy 
$E=E_2$ in the case of the repulsive Coulomb interaction ($q_1 q_2 >0$) }\\ 

Taking the integration over $\omega$ in the expression (27 we obtain simple analytical expressions 
for the partial wave Coulomb $t$-matrix at $E=E_2$ in the case of like charges ($\gamma=2$). In 
particular, the expression for the partial $s$-wave Coulomb t-matrix in the case of the repulsive 
interaction has the form
\begin{displaymath}
t_0^r(k, k^{\prime}; E_2) = \frac{\pi q_1 q_2}{k k^{\prime}} \left\{ 16 (3\xi - 2)\eta - 
8(2\xi - 1)\eta \rho_2  + (8 \xi^2 - 8\xi + 8 \eta^2 + 1) \ln \frac{\xi + \eta}{\xi - \eta} \right.
\end{displaymath}
\begin{equation}
 \left. + 8 (2\xi - 1)\eta \ln (\xi^2 - \eta^2) - (\pi - \omega_{\pi})\sin 2\omega_{\pi} 
+ (\pi - \omega_0)\sin 2\omega_0 \right\}\;.  
\end{equation}   

Similarly we get the analytical expressions for the partial $p$- and $d$- wave Coulomb $t$-matrices: 
\begin{displaymath}
t_1^r(k, k^{\prime}; E_2) = \frac{\pi q_1 q_2}{k k^{\prime}} \left\{ (\frac{8}{3}\xi -16 \eta^2 - 2) + 
\frac{16}{3}\eta^2 \rho_2 \right.
\end{displaymath}
\begin{displaymath}
+ \frac{1}{\eta} \left( \frac{8}{3} \xi^3 - 4\xi^2 + \xi - 8 \xi\eta^2 + 4\eta^2\right) 
\ln \frac{\xi + \eta}{\xi - \eta} 
-\frac{16}{3} \eta^2 \ln \left( \xi^2 - \eta^2\right) 
\end{displaymath}
\begin{equation}
-\frac{1}{2\eta}\left[\left(\pi - \omega_{\pi}\right)\sin \omega_{\pi} 
- \left( \pi - \omega_0 \right)\sin \omega_0 \right]
\end{equation}  
\begin{displaymath}
-\frac{1}{2\eta}(2\xi - 1) \left[\left(\pi - \omega_{\pi}\right)\sin 2\omega_{\pi} 
- \left( \pi - \omega_0 \right)\sin 2\omega_0 \right] 
\end{displaymath}
\begin{displaymath}
\left.-\frac{1}{6\eta} \left[\left(\pi - \omega_{\pi}\right)\sin 3\omega_{\pi} 
- \left( \pi - \omega_0 \right)\sin 3\omega_0 \right]\right\}   
\end{displaymath} 
and 
\begin{displaymath}
t_2^r(k, k^{\prime}; E_2) = \frac{\pi q_1 q_2}{k k^{\prime}} \left\{ \frac{1}{\eta}\left( 2\xi^2 
- \frac{5}{2} \xi - 2 \eta^2 + \frac{3}{4} \right) \right.
\end{displaymath} 
\begin{displaymath}
+  \frac{1}{\eta^2} \left[ \left( -2\xi^4 
+4\xi^3 + \frac{3}{2}\xi^2 + 4\xi^2\eta^2 + 12\xi\eta^2 - \frac{1}{2}\eta^2) -2\eta^4 \right) 
\ln \frac{\xi + \eta}{\xi - \eta}\right. 
\end{displaymath}
\begin{equation}
 -\frac{3\pi}{16}\left( \omega_{\pi} - \omega_{0}\right)
 + \frac{3}{32}\left( \omega_{\pi}^2 - \omega_{0}^2\right)
-\frac{3}{4}(2\xi - 1) \left[\left(\pi - \omega_{\pi}\right)\sin \omega_{\pi} 
- \left( \pi - \omega_0 \right)\sin \omega_0 \right]
\end{equation} 
\begin{displaymath}
-\left(\frac{3}{2}\xi^2-\frac{3}{2}\xi-
\frac{1}{2}\eta^2+\frac{9}{16}\right) \left[\left(\pi - \omega_{\pi}\right)\sin 2\omega_{\pi} 
- \left( \pi - \omega_0 \right)\sin 2\omega_0 \right]
\end{displaymath} 
\begin{displaymath}
-\frac{1}{4}(2\xi - 1)\left[\left(\pi - \omega_{\pi}\right)\sin 3\omega_{\pi} 
- \left( \pi - \omega_0 \right)\sin 3\omega_0 \right] 
\end{displaymath} 
\begin{displaymath}
\left.\left. -\frac{3}{64}\left[\left(\pi - \omega_{\pi}\right)\sin 4\omega_{\pi} 
- \left( \pi - \omega_0 \right)\sin 4\omega_0 \right] \right] \right\}\;.
\end{displaymath}  

Note that the formulae for $\xi$ and $\eta$ (20), as well as the formulae for $\omega_0$ and 
$\omega_{\pi}$ (22) contain $\kappa = \kappa_2$, corresponding to the energy $E_2$ (7).\\
 
\noindent {\sf 6. Analytical expressions for the partial-wave transition matrix at 
the energy $E=E_2$ in the case of the attractive Coulomb interaction ($q_1 q_2 <0$)}\\ 

As is started at the end of Section 4, according to (31), the partial-wave components of the 
transition matrices for the particles with the unlike charges $t_l^a(k, k^{\prime}; E)$ with 
$l=0$ and $l=1$ are singular at the point $E=E_2.$ The lowest nonsingular at this point 
partial-wave component is the $d$-wave component with $l=2$. Taking the corresponding integration 
over $\omega$ in the expression (29) $l=2$ we obtain the following formula for  
$t_2^a(k, k^{\prime}; E_2)$:
\begin{displaymath}
t_2^a(k, k^{\prime}; E_2) = \frac{\pi q_1 q_2}{k k^{\prime}} \left\{ \frac{1}{\eta}\left( -8\xi^3 
+ 14 \xi^2 -\frac{7}{2}\xi +\frac{40}{3}\xi\eta^2 - \frac{26}{3}\eta^2 - \frac{3}{4}\right) \right.
\end{displaymath} 
\begin{displaymath}
+  \frac{1}{\eta^2} \left[ \left( -2\xi^4 
+4\xi^3 + \frac{3}{2}\xi^2 + 4\xi^2\eta^2 + 12\xi\eta^2 - \frac{1}{2}\eta^2) -2\eta^4 \right) 
\ln \frac{\xi + \eta}{\xi - \eta}\right. 
\end{displaymath}
\begin{equation}
+\frac{3\pi}{16}\left( \omega_{\pi} - \omega_{0}\right)
- \frac{3}{32}\left( \omega_{\pi}^2 - \omega_{0}^2\right)
+\frac{3}{4}(2\xi - 1) \left[\left(\pi - \omega_{\pi}\right)\sin \omega_{\pi} 
- \left( \pi - \omega_0 \right)\sin \omega_0 \right]
\end{equation} 
\begin{displaymath}
+\left(\frac{3}{2}\xi^2-\frac{3}{2}\xi-
\frac{1}{2}\eta^2+\frac{9}{16}\right) \left[\left(\pi - \omega_{\pi}\right)\sin 2\omega_{\pi} 
- \left( \pi - \omega_0 \right)\sin 2\omega_0 \right]
\end{displaymath} 
\begin{displaymath}
+\frac{1}{4}(2\xi - 1)\left[\left(\pi - \omega_{\pi}\right)\sin 3\omega_{\pi} 
- \left( \pi - \omega_0 \right)\sin 3\omega_0 \right]
\end{displaymath} 
\begin{displaymath}
\left.\left. +\frac{3}{64}\left[\left(\pi - \omega_{\pi}\right)\sin 4\omega_{\pi} 
- \left( \pi - \omega_0 \right)\sin 4\omega_0 \right] \right] \right\}\;.
\end{displaymath}  
It is interesting to note, with the difference in sign for $q_1 q_2 $ in the coefficients before 
the braces in the expressions (34) and (35), that the formulae for the corresponding 
partial $d$-wave transition matrices in the cases of the repulsive and attractive Coulomb 
interactions differ only by their first terms and the signs in front the terms with the logarithm. 
The remaining terms are the same. \\
 
\noindent {\sf 7. Discussion and conclusions}\\ 

The investigation of the properties of the off-energy-shell two-body partial-wave Coulomb 
transition matrices is necessary in connection with the formulation and solution of the integral 
equations for composite atomic and nuclear systems.

The possibility of the analytical derivation of the  two-body partial-wave Coulomb transition 
matrices with the use of the specific symmetry of the Coulomb system in Fock's four-dimensional 
Euclidean space has been first studied in our preceding paper [20]. Therein the simple analytical 
expressions for the partial $p$-, $d$- and $f$-wave Coulomb transition matrices at the energy of 
the qround bound state $E=E_1$ (that corresponds to the Coulomb parameter $\gamma=-1$) has been 
derived in the case of the oppositely charged particles. In the case of the repulsive Coulomb 
potential analytical expressions for the partial $s$-, $p$- and $d$-wave transition matrices 
at the energy $E=E_1$ have been obtained in Ref. [21]. 

The knowledge of partial-wave Coulomb transition matrix is necessary to determine the electric
$2^{\lambda}$-pole polarizabilities of the two-particle Coulomb system 
$\alpha_{\lambda}$ ($\lambda=1,2\cdots$) in the  state with the energy $E_n$ [23,24]. 

In this paper the Fock's method has been applied to derive the partial-wave Coulomb transition 
matrices at the first excited state energy $E=E_2$. The cases of the two-body system both with  
repulsive and attractive Coulomb interactions have been investigated (the transition matrices 
$t_l^r(k, k^{\prime}; E_2)$ and  $t_l^a(k, k^{\prime}; E_2)$). It is interesting that the simplification 
of the analytical form of the partial-wave transition matrix at the energies of the bound states 
takes place both for the particles with opposite charges and for the particles with like-sign 
charges bound states of which are not formed at all.

At present it is essentially interesting to study the possibility of derivation of expressions for 
the partial-wave transition matrix on the basis of the one-parameter integral representation of the 
three-dimensional Coulomb transition matrix, proposed by Schwinger [13]. Also, of special importance 
is the analytical solution of the integral Lippmann-Schwinger equation for the partial-wave Coulomb 
transition matrices with the values of the energy $E$ differing from the energies of both the ground 
and excited bound states, for example, in the case of the half-integer values of the Sommerfeld 
parameter, $\gamma=+1/2$ and $\gamma=-1/2$, when all the partial-wave components are singularity-free 
in the energy. \\ 

\noindent {\sf Acknowledgment}\\ 

The present work was partially supported by the National Academy of Sciences 
of Ukraine (project No. 0117U00237) and by the Program of Fundamental Research 
of the Department of Physics and Astronomy of NASU (project No. 0117U00240). 
\\ [.2in] 

\noindent {\footnotesize {\sf References} 
\vspace*{.1in}
\begin{itemize} 
\setlength{\baselineskip}{.1in} 
\item[{\tt [1]}] Chen J C Y, Chen A C 1972 Advances in Atomic and Molecular Physics 
                  ed D B Bates and I Estermann vol 8 (N Y - London: Academic Press)pp 71-129
\item[{\tt [2]}] Faddeev L D 1961 Sov. Phys. JETP 12 1014-9 
\item[{\tt [3]}] Faddeev L D 1965 Mathematical Aspects of the Three-Body Problem in the Quantum
                 Scattering Yheory (Isr Program Sci Transl, Jerusalem: 1965) 
\item[{\tt [4]}] Yakubovsky O A 1967 Sov. J. Nucl. Phys. 5 937-42    
\item[{\tt [5]}] Vesselova A M 1970 Teor. Mat. Fiz. 3 326-31
\item[{\tt [6]}] Vesselova A M 1972 Teor. Mat. Fiz. 13 368-76  
\item[{\tt [7]}] Gorshkov V G 1961 Zh. Eksp. Teor. Fiz. 40 1481-90               
\item[{\tt [8]}] Okubo S and Feldman D 1960 Phys. Rev. 117 292-306
\item[{\tt [10]}] Bratsev V F and Trifonov  E D 1962 Vest. Leningrad. Gos. Univ. 16 36-9 
\item[{\tt [11]}] Hostler L 1964 J. Math. Phys. 5 591-611
\item[{\tt [12]}] Hostler L 1964 J. Math. Phys. 5 1235-40
\item[{\tt [13]}] Schwinger J 1964 J. Math. Phys. 5 1606-8.
\item[{\tt [14]}] Perelomov A M and Popov V S 1966 Sov. Phys. JETP 23 118-34  
\item[{\tt [15]}] Shadchin S A and Kharchenko V F 1983 J. Phys. B: At. Mol.Phys. 16 1319-22 
\item[{\tt [16]}] Storozhenko S A and Shadchin S A 1988 Teor. Mat. Fiz. 76 339-49 
\item[{\tt [17]}] Kok L P and van Haeringen H Phys. Rev. 1980 C 21 512-7
\item[{\tt [18]}] van Haeringen H 1984 J. Math. Phys. 25 3001-32
\item[{\tt [19]}] Fock V A 1935 Z. Phys. 98 145-54 
\item[{\tt [20]}] Kharchenko V F 2016 Ann. Phys. NY 374 16-26
\item[{\tt [21]}] Kharchenko V F 2017 Ukr. J. Phys. 62 263-70
\item[{\tt [22]}] Gradstein I S and Ryzhik I M 1971 Tables of Integrals, Sums,Series and Products 
                  (Moscow: Nauka)
\item[{\tt [23]}] Kharchenko V F 2013 J. Mod. Phys. 4 99-107
\item[{\tt [24]}] Kharchenko V F 2015 Ann. Phys. NY 355 153-69

\end{itemize}} 

\end{document}